\pdfoutput=1
\documentclass[
reprint,prx, longbibliography,
superscriptaddress,
 amsmath,amssymb,
 aps,
]{revtex4-2}

\usepackage{graphicx}
\usepackage{dcolumn}
\usepackage{bm}
\usepackage{siunitx}
\usepackage{chemformula}
\usepackage{CJKutf8}

\newcommand{\ket}[1]{\left| #1 \right>} 
\newcommand{\bra}[1]{\left< #1 \right|} 

\newboolean{ShowComments}
\setboolean{ShowComments}{true}  
\ifthenelse{\boolean{ShowComments}}%
	{
		\newcommand{\ColorComment}[3]{%
				{\colorbox{#1}{\color{white}   \textsf{\textbf{#2}}} \textcolor{#1}{#3}}}
		\newcommand{\nyacite}[1]{[#1]}
	}%
	{
		\newcommand{\ColorComment}[3]{}
		\newcommand{\nyacite}[1]{}
	}%
\definecolor{jcmcolor}{rgb}{0,0.5,0}
\definecolor{fjhcolor}{rgb}{0,0.3,1}
\definecolor{tbdcolor}{rgb}{0.5,0,0}

\begin{document}
\begin{CJK}{UTF8}{gbsn}
\preprint{APS/123-QED}

\title{Guiding Diamond Spin Qubit Growth with Computational Methods}

\author{Jonathan C. Marcks}
\thanks{These two authors contributed equally}
\affiliation{Pritzker School of Molecular Engineering, University of Chicago, Chicago, IL, 60637, United States}
\affiliation{Center for Molecular Engineering and Materials Science Division, Argonne National Laboratory, Lemont, IL, 60439, United States}

\author{Mykyta Onizhuk}
\thanks{These two authors contributed equally}
\affiliation{Department of Chemistry, University of Chicago, Chicago, IL, 60637, United States}
\affiliation{Pritzker School of Molecular Engineering, University of Chicago, Chicago, IL, 60637, United States}

\author{Nazar Delegan}
\affiliation{Center for Molecular Engineering and Materials Science Division, Argonne National Laboratory, Lemont, IL, 60439, United States}
\affiliation{Pritzker School of Molecular Engineering, University of Chicago, Chicago, IL, 60637, United States}

\author{Yu-Xin Wang (王语馨)}
\affiliation{Pritzker School of Molecular Engineering, University of Chicago, Chicago, IL, 60637, United States}

\author{Masaya Fukami}
\affiliation{Pritzker School of Molecular Engineering, University of Chicago, Chicago, IL, 60637, United States}

\author{Maya Watts}
\affiliation{Center for Molecular Engineering and Materials Science Division, Argonne National Laboratory, Lemont, IL, 60439, United States}
\affiliation{Pritzker School of Molecular Engineering, University of Chicago, Chicago, IL, 60637, United States}

\author{Aashish A. Clerk}
\affiliation{Pritzker School of Molecular Engineering, University of Chicago, Chicago, IL, 60637, United States}

\author{F. Joseph Heremans}
\affiliation{Center for Molecular Engineering and Materials Science Division, Argonne National Laboratory, Lemont, IL, 60439, United States}
\affiliation{Pritzker School of Molecular Engineering, University of Chicago, Chicago, IL, 60637, United States}

\author{Giulia Galli}
\affiliation{Pritzker School of Molecular Engineering, University of Chicago, Chicago, IL, 60637, United States}
\affiliation{Department of Chemistry, University of Chicago, Chicago, IL, 60637, United States}
\affiliation{Center for Molecular Engineering and Materials Science Division, Argonne National Laboratory, Lemont, IL, 60439, United States}

\author{David D. Awschalom}
\affiliation{Pritzker School of Molecular Engineering, University of Chicago, Chicago, IL, 60637, United States}
\affiliation{Department of Physics, University of Chicago, Chicago, IL, 60637, United States}
\affiliation{Center for Molecular Engineering and Materials Science Division, Argonne National Laboratory, Lemont, IL, 60439, United States}

\date{\today}

\begin{abstract}
The nitrogen vacancy (NV) center in diamond, a well-studied, optically active spin defect, is the prototypical system in many state-of-the-art quantum sensing and communication applications. In addition to the enticing properties intrinsic to the NV center, its diamond host's nuclear and electronic spin baths can be leveraged as resources for quantum information, rather than considered solely as sources of decoherence. However, current synthesis approaches result in stochastic defect spin positions, reducing the technology's potential for deterministic control and yield of NV-spin bath systems, as well as scalability and integration with other technologies. Here, we demonstrate the use of theoretical calculations of electronic central spin decoherence as an integral part of an NV-spin bath synthesis workflow, providing a path forward for the quantitative design of NV center-based quantum sensing systems. We use computationally generated coherence data to characterize the properties of single NV center qubits across relevant growth parameters to find general trends in coherence time distributions dependent on spin bath dimensionality and density. We then build a maximum likelihood estimator with our theoretical model, enabling the characterization of a test sample through NV $T_2^*$ measurements. Finally, we explore the impact of dimensionality on the yield of strongly coupled electron spin systems. The methods presented herein are general and applicable to other qubit platforms that can be appropriately simulated.

\end{abstract}

\maketitle
\end{CJK}

\section{\label{sec:introduction}Introduction}
Defect color centers in diamond~\cite{Doherty2013,Bradac2019} have been demonstrated as quantum magnetometers~\cite{Balasubramanian2008,Maze2008,Maertz2010,LeSage2013,Rondin2014,Mamin2013,Staudacher2013,Casola2018,Zhao2022} and nodes in quantum communication networks~\cite{Kalb2017,nguyen2019a,nguyen2019b,Pompili2021,Hermans2022}. Quantum applications of the negatively charged nitrogen vacancy (NV) center, with a spin-photon interface and coherent operation up to and above room temperature~\cite{Toyli2012,Liu2019,Doherty2013,Heremans2016}, will benefit from interfacing the central NV spin qubit with accessible dark spins in the diamond lattice for quantum memories~\cite{Pfender2017,Bradley2019,Degen2021} and many-body metrological states~\cite{Xie2021,Zheng2022,Meriles2023}. These applications could enable scalable quantum networks and quantum sensing beyond the standard quantum limit. Explorations of such multi-spin systems have relied on NV centers that are either naturally occurring~\cite{Hanson2006,Knowles2016,Bradley2019,Degen2021,Pompili2021,Hermans2022}, precluding scalability, or that are formed via nitrogen implantation~\cite{Meijer2005,Rabeau2006,Gaebel2006,Neumann2010,Dolde2013,Rosenfeld2018,Cooper2020,Lee2022}, introducing qubit decoherence sources, associated with crystal damage~\cite{Yamamoto2013}.

Diamond-based quantum applications benefit greatly from the ongoing optimization of bottom-up color center synthesis via plasma-enhanced chemical vapor deposition (PECVD)~\cite{Okushi2002,Ohno2014b,Eichhorn2019,Balasubramanian2022}. Delta($\delta$)-doping studies~\cite{Ohno2012,Ohno2014,McLellan2016} have demonstrated vacancy diffusion-limited spatially localized NV centers, while avoiding the crystal damage and processing inherent to aperture mask or focused implantation~\cite{Toyli2010,Spinicelli2011,Ohno2014,Sangtawesin2014,Jakobi2016,Bayn2015,Hwang2022}. PECVD of diamond quantum systems has enabled engineering of NV center spin environments via isotopic purification~\cite{Mizuochi2009,Balasubramanian2009,Ohno2012}, dimensionality control~\cite{Ohno2012,Davis2023,Hughes2023}, and co-doping techniques~\cite{luhmann2019,Herbschleb2019,Kawase2022}. However, the development of these techniques has outpaced computational efforts to model spin bath-induced decoherence~\cite{Bauch2020,Park2022}, and theoretical approaches have not yet been applied to investigate diamond qubit synthesis. Cluster Correlation Expansion (CCE) methods provide an accurate approach to model decoherence in varied and tailored electron and nuclear spin bath environments~\cite{Onizhuk2021a}. Such methods have recently been applied to study material systems relevant for quantum applications~\cite{Onizhuk2021b,Kanai2022,Bayliss2022,Park2022}, indicating that CCE may indeed be a powerful tool to enable more efficient synthesis procedures, which are crucial for the design of quantum materials~\cite{deleon2021}.

In this work, we apply CCE methods, as implemented in the open source framework PyCCE~\cite{Onizhuk2021b}, to predict and characterize bottom-up solid state spin qubit synthesis. We first introduce the computation and materials growth techniques. We then explore a common electronic spin defect created during NV center synthesis: the neutrally charged substitutional nitrogen \ch{N_s^0} with electron spin $S=1/2$ (P1 center). Using theoretical predictions, we investigate the P1 center electron spin bath-induced decoherence~\cite{Bauch2020,Shinei2022} of NV centers in diamond across the parameter space of our growth regime (P1 density and layer thickness). We focus on low-dimensional spin bath geometries, finding central spin lifetime-limited coherence times and a non-trivial dependence of single spin coherence times on dimensionality. Obtained dependencies enable the use of coherence time distributions as descriptors of these systems for determining the growth parameters. To this end, we develop a maximum likelihood estimation (MLE) model based on Ramsey $T_2^*$ coherence times and apply it to characterize nitrogen incorporation in a experimental test sample. We then study low-dimensional electron spin baths as hosts to strongly coupled electron spin systems, demonstrating how our computational techniques can help improve the yield of future quantum devices and aid in experimental design.

\begin{figure}
    \centering
    \includegraphics{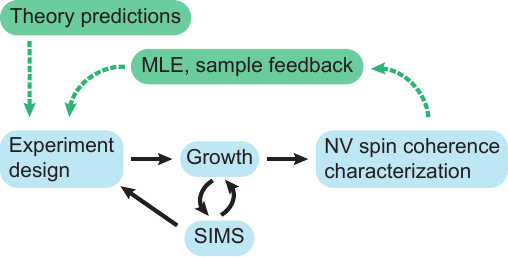}
    \caption{\bf Growth process workflow. \rm The current process steps (blue) for synthesizing a diamond NV sample. Iterations of growth and SIMS analysis are required to confirm nitrogen doping densities. The theoretical predictions and density maximum likelihood estimation model in this work (green) enable a non-destructive feedback process to circumvent SIMS and allow for an efficient experimental design.}
    \label{fig:fig1}
\end{figure}

In Fig. \ref{fig:fig1} we show the strategy adopted in this work to improve upon the current NV synthesis process. The blue boxes show the commonly adopted process for generating single NV centers. After identifying a desired sample density and geometry, iterations of growth and secondary ion mass spectroscopy (SIMS) are necessary to confirm the nitrogen doping density. In practice, we have observed large variations in SIMS results that reduce the efficacy of this approach, as discussed in Sec. \ref{subsec:growth}. Here we show that it is beneficial to incorporate theoretical spin bath predictions as well as an \emph{in-situ} density characterization tool into our workflow (green boxes). The understanding of low-dimensional spin bath decoherence obtained through theory and computation improves initial experiment design, and the local density feedback enabled by the MLE model circumvents the need for SIMS characterizations of doping density.

\begin{figure}
    \centering
    \includegraphics{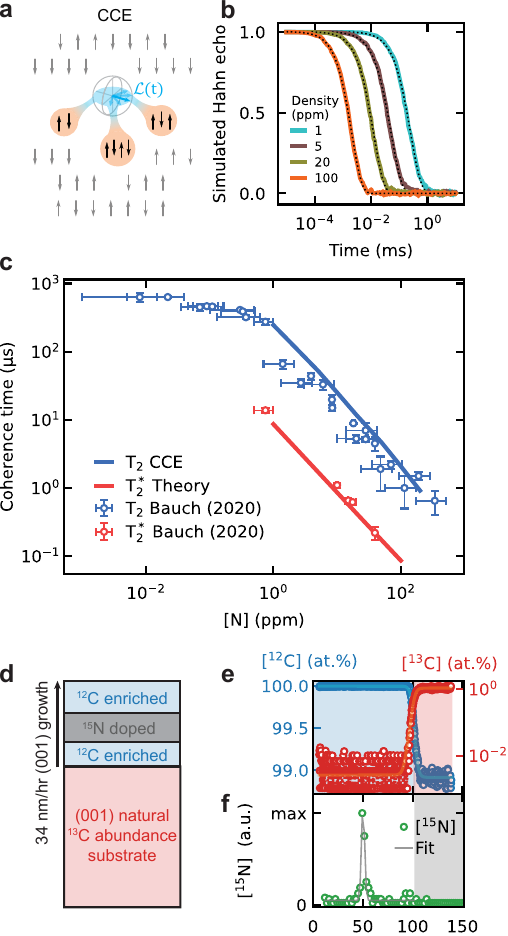}
    \caption{\bf Computational and diamond growth methods. \rm (a)  Schematic representation of the cluster correlation expansion (CCE) approach. (b) Example of the Hahn-echo coherence calculated using the PyCCE code~\cite{Onizhuk2021b} for various $\ch{^{14}N}$ P1 spin baths. The values of $T_2$ times are extracted from a stretched exponential fit of the form $\exp{[-(\frac{t}{T_2})^n]}$ (dashed line). (c) $T_2$ and $T_2^*$ coherence times overlaid with corresponding experimental data~\cite{Bauch2020}, validating our computational methods. (d) Schematic of isotopically pure ($^{12}$C) PE-CVD (100) diamond overgrowth with isotopically tagged $^{15}$N nitrogen $\delta$-doping. This sample geometry with varying nitrogen incorporation density and thickness is considered throughout this paper. (e,f) Carbon (top) and nitrogen (bottom) isotope concentrations measured via SIMS on characterization sample, demonstrating isotopic purification of host material and isotopically tagged nitrogen incorporation. Carbon SIMS is used to calibrate growth rate, shown in (d). The nitrogen concentration is quantified with NV coherence measurements in Sec.~\ref{subsec:mle}.}
    \label{fig:fig2}
\end{figure}

\section{\label{sec:results}Results}

\subsection{\label{subsec:techniques}Validation of theoretical calculations}
Within the CCE approach~\cite{Yang2008,Yang2009} the coherence function $\mathcal{L}(t) = \frac{\bra{0}\hat \rho (t) \ket{1}}{\bra{0}\hat \rho (0) \ket{1}}$, defined as the normalized off-diagonal element of the density matrix of the qubit $\hat \rho (t)$, is approximated as a product of irreducible contributions of bath spin clusters, where the maximum size of the cluster $n$ corresponds to the order $n$ of the CCE$n$ approximation (Fig.~\ref{fig:fig2}(a)). We converge the calculations with respect to the size of the bath, and the order of CCE approximation. We find that the Ramsey signal of the electron spin in the electron spin bath is converged at first order (CCE1), when each P1 is treated as isolated spin. We can thus solve the P1-limited Ramsey decoherence analytically, and compute $T_2^*$ as a sum of the couplings between the NV center and the weakly coupled P1 centers. The Hahn echo signal is instead simulated at the CCE4 level of theory (see Methods for more details).

We validate our theoretical calculations against a reference dataset of NV center ensemble coherence times in bulk \ch{^{14}N} P1 spin baths~\cite{Bauch2020}. We extract $T_2$ from the coherence curve by fitting the signal to a stretched exponential function, $\exp{[-(\frac{t}{T_2})^n]}$, as shown in Fig.~\ref{fig:fig2}(b). Computed ensemble $T_2^*$ and $T_2$, averaged over a set of random P1 positions, are overlaid in Fig.~\ref{fig:fig2}(c) with experimental data, taken from Ref.~\cite{Bauch2020}. We find excellent agreement with the experimental data, showing that the first-principles calculation with CCE method yields a quantitative description of the decoherence due to P1 spin baths. The stretched exponent parameter of the computed Hahn-echo decay is between $n=1.2\text{-}1.3$, in excellent agreement with the data of Ref.~\cite{Bauch2020}.

\subsection{\label{subsec:growth}Diamond growth and defect synthesis}

The sample studied in this paper, shown schematically in Fig.~\ref{fig:fig2}(d) was grown with a \SI{3}{\minute}, \SI{10}{sccm} \ch{^{15}N2} flow at a time corresponding to a depth of \SI{\approx50}{\nano\meter}. Nitrogen $\delta$-doping is achieved by introducing \ch{^{15}N2} gas (\SI{99.99}{\percent} chemical purity, \SI{99.9}{at\percent} isotopic purity) during diamond growth. According to the SIMS characterization of a calibration sample, shown in Fig.~\ref{fig:fig2}(e-f), this creates a \SI{3.8\pm0.2}{\nano\meter} thick (compared to $1.3^{+2.2}_{-0}$\SI{}{\nano\meter} predicted from growth calibrations) \ch{^{15}N}-doped layer at a depth of \SI{50.2\pm0.1}{\nano\meter}, with a SIMS-quantified \ch{[^{15}N]} density of \SI{0.39\pm0.02}{ppm} within \ch{^{12}C} isotopically purified diamond overgrowth ([\ch{^{12}C}]=\SI{99.993}{at\percent}). These values are obtained from a calibration sample, processed and grown identically to the sample studied in this paper.

While SIMS is often used for detecting low concentrations of dopants in semiconducting materials, sample geometries unique to our application remain difficult to characterize accurately due to experimental limitations. Specifically, the trade-off between depth resolution and overall sensitivity is dictated by the analysis/sputtering energy. Under our characterization conditions, the ideal detection limits for \ch{^{15}N2} and \ch{^{14}N2} densities are \SI{1e15}{\centi\meter^{-3}} (\SI{\approx  0.006}{ppm}) and \SI{5e15}{\centi\meter^{-3}} (\SI{\approx 0.028}{ppm}), respectively. However, the obtained densities can vary significantly as a function of sample inhomogeneities, the presence of growth defects, and experimental conditions. While studying samples that were nominally grown under the same conditions, SIMS quantification of \ch{[^{15}N]} has been observed to regularly vary by at least an order of magnitude, requiring rigorous statistics over growth of multiple samples, a time- and resource-consuming process. A truly local spin-defect materials characterization method is necessary, motivating the \emph{in-situ} maximum likelihood estimation of the density characterization presented in Sec.~\ref{subsec:mle}, a new capability enabled by our computational results. A different approach with NV ensemble coherence measurements has also recently been developed~\cite{Davis2023,Hughes2023}.

\begin{figure*}
    \centering
    \includegraphics[width=172mm]{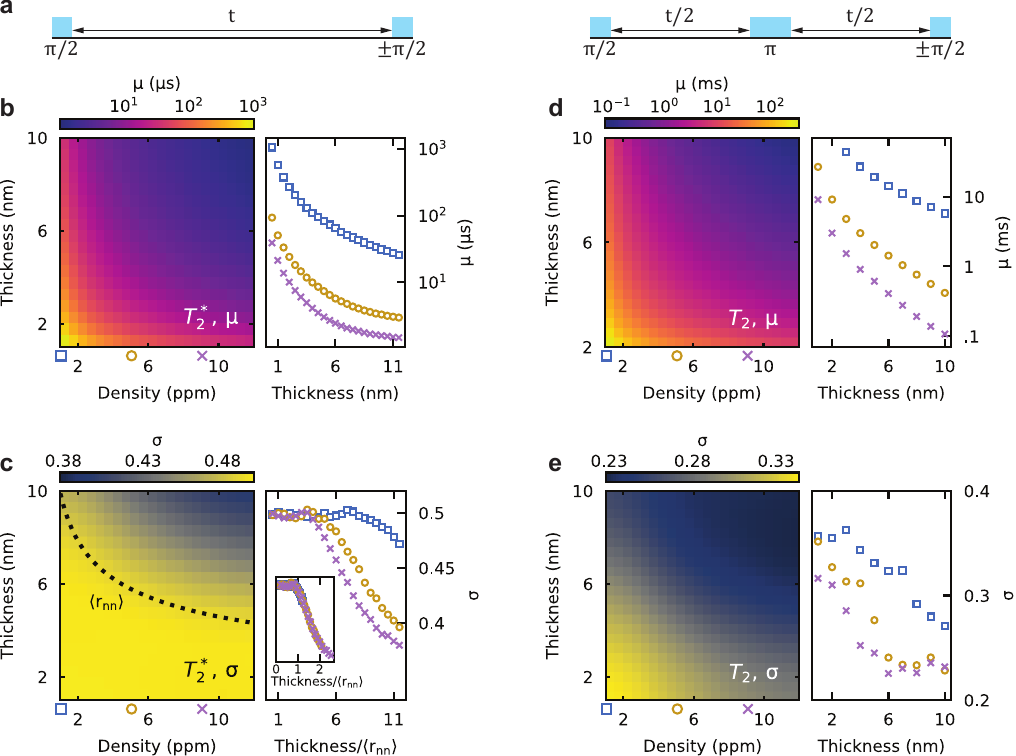}
    \caption{\bf Single spin coherence in low-dimensional spin baths.  \rm (a) Ramsey (left) and Hahn echo (right) microwave measurement pulse sequences. (b,c) Mean of $\log_{10} (T_2^*/[\SI{1}{\milli\second}])$ distributions $\mu=10^{\langle\log_{10} (T_2^*/[\SI{1}{\milli\second}])\rangle}$ (b) and variance $\sigma^2=\langle\log_{10}^2 (T_2^*/[\SI{1}{\milli\second}])\rangle-\langle\log_{10} (T_2^*/[\SI{1}{\milli\second}])\rangle^2$ (c) as a function of P1 density and layer thickness. Values are linearly interpolated between datapoints. The black dashed line in (c) indicates the thickness equal to the average nearest-neighbor bath spin distance $\left<r_{nn}\right>=0.554\rho^{-1/3}$ for each density $\rho$ (see text, Sec.~\ref{sec:NVP1}), demonstrating a boundary between dimensionalities. At right in (b,c) are line-cuts of $\mu$ and $\sigma$ at densities of 1, 5, and \SI{9}{ppm}. Inset in (c) is $\sigma$ at multiple densities with thickness normalized by $\left<r_{nn}\right>$, demonstrating universal behavior versus dimensionality. (d,e) Same data as (b,c) presented for $T_2$ coherence times.}
    \label{fig:fig3}
\end{figure*}

\subsection{\label{subsec:singlespin}Single spin coherence in quasi-2D electron bath}
We investigate single spin coherence properties across the density and thickness parameter space available for the PECVD growth recipe adopted in this work and described in Methods section. We compute Ramsey coherence time $T_2^*$ (Fig.~\ref{fig:fig3}(a), left) for $10^4$ spin bath configurations with spin bath thickness of \qtyrange{0.5}{12}{\nano\meter} (\SI{0.5}{\nano\meter} steps) and density of \qtyrange{0.5}{12}{ppm} (\SI{0.5}{ppm} steps) from the coupling between the central NV center electron spin and weakly coupled P1 center electron spins (see Methods). We simulate Hahn-echo measurements (Fig.~\ref{fig:fig3}(a), right) with spin bath thicknesses of \qtyrange{1}{10}{\nano\meter} (\SI{1}{\nano\meter} steps) and densities of \qtyrange{1}{12}{ppm} (\SI{1}{ppm} steps) (See SI Sec.~IB for justification of CCE order).

We characterize the distributions of the coherence times with the mean $\mu=10^{\langle\log_{10} (T_2/[\SI{1}{\milli\second}])\rangle}$ and the variance $\sigma^2=\langle\log_{10}^2 (T_2/[\SI{1}{\milli\second}])\rangle-\langle\log_{10} (T_2/[\SI{1}{\milli\second}])\rangle^2$ of the logarithm of the coherence times at each density and thickness (Fig.~\ref{fig:fig3}). Using the logarithm of the coherence we can directly compare the coherence distributions at different timescales.

Figs. \ref{fig:fig3}(b,c) and (d,e) depict $\mu$ and $\sigma$ over the chosen range of parameters for $T_2^*$ and $T_2$, respectively. In each case, the computed average coherence time decreases with increasing spin density and/or increasing thickness, as expected. In the three-dimensional limit, the average coherence time is independent of bath thickness. The observed decrease in $\mu$ as a function of thickness (Fig.~\ref{fig:fig3}(b) and (d)) suggests the presence a low-dimensional spin bath regime in the chosen range of parameters.

We analytically derive the distribution of the interaction strength between the central spin and bath spins in low-dimensional baths in Sec.~\ref{sec:NVP1}. In the case of $T_2$ we predict times $>$\SI{1}{\milli\second} in the bottom-left half of the parameter space in Fig.~\ref{fig:fig3}(d), beyond what is generally observed in experiment. This suggests that experimental $T_2$ times in thin, low density spin baths are limited by noise sources not captured in our model, as suggested previously~\cite{Bauch2020}. However, our calculations predict that, in principle, low-dimensional lightly doped samples can realize $T_1$ limited coherence times at room temperature.

Bath dimensionality further impacts the relative distribution of coherence times, described by the standard deviation $\sigma$. Focusing on the inhomogeneous dephasing time $T_2^*$ (Fig.~\ref{fig:fig3}(c), right), $\sigma$ exhibits unexpected behavior in the region where the thickness equals the average nearest neighbor distance in three dimensions, $\left< r_{nn}\right>$, plotted as a function of density in the left plot. $\sigma$ plateaus when the thickness is smaller than $\left< r_{nn}\right>$ and decreases when thicknesses are larger. The inset in Fig.~\ref{fig:fig3}(c), right, demonstrates universal behavior of coherence times relative to the bath dimensionality. The $x$-axis is normalized to $\left< r_{nn}\right>$. This indicates that two-dimensional spin baths naturally have a wider spread of NV center coherence times. While thin and less dense samples may optimize coherence times, they typically also lead to greater fluctuations in single-qubit coherence properties.

We see similar trends in Hahn-echo $T_2$ times (Fig.~\ref{fig:fig3}(e), right). We find in general that $\sigma_{T_2^*}>\sigma_{T_2}$. In the SI Sec.~IC, we find convergence for $T_2^*$ and $T_2$ at twelve and 100 bath spins, respectively, suggesting heuristically that Ramsey measurements are sensitive to the variation of a fewer number of spins. In general, one expects a smaller standard deviation in physical quantities that are sensitive to larger numbers of randomly placed spins due to the central limit theorem. We thus expect a larger impact of the stochasticity in P1 position on the $T_2^*$ distributions. These results inform solid-state qubit synthesis characterization, where both $T_2^*$ and $T_2$ are standard measurements performed on multiple NV centers.

Our theoretical results constitute a full computational characterization of spin-bath induced coherence times across a range of bath geometries and densities. Our computational strategy is not limited to NV centers in diamond and can be applied to other spin defect systems, as well as other spin bath measurements, as long as the appropriate pulse sequence can be simulated using the PyCCE code. Additionally, our approach will inform future diamond growth and NV synthesis. Rather than extrapolating from bulk data~\cite{Bauch2020} or measurements on single $\delta$-doped NV centers, growth may now be informed by the theoretical predictions of coherence times distributions.

\subsection{\label{sec:coherence}Sample characterization}
\begin{figure}
    \centering
    \includegraphics[width=86mm]{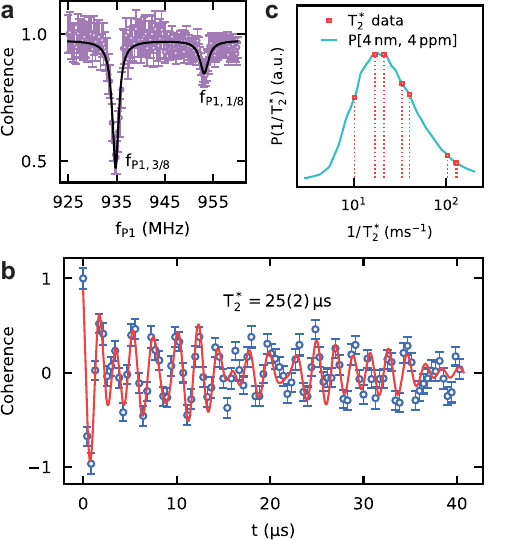}
    \caption{\bf NV center measurements. \rm (a) DEER spectroscopy with NV center confirming the presence of a P1 center electron spin bath. Marked values of $f_{P1}$ correspond to P1 ESR transitions corresponding to the static magnetic field of \SI{311}{G} and internal P1 hyperfine parameters. (b) Ramsey interferometry measurement to extract $T_2^*$ coherence time. (c) Compiled decoherence rates for eight measured NV centers (the two right-most points overlap) overlaid with the probability function $P$, drawn directly from the histogram of computed coherence times for parameters determined in Sec.~\ref{subsec:mle}. The height of each data point indicates the calculated probability for each value for that time, and does not represent an experimental value.}
    \label{fig:fig4}
\end{figure}

We characterize the coherence of an exemplar sample grown under the conditions outlined in Sec.~\ref{subsec:methods:growth}.
Fig.~\ref{fig:fig4}(a) presents frequency-dependent double electron-electron resonance (DEER) measurements of a single NV center in a P1 center bath. This measurement essentially performs electron spin resonance (ESR) spectroscopy on target spins (here P1 center electron spins) by recoupling their dipolar interactions to the NV center probe spin, which are otherwise decoupled by the Hahn-echo sequence. At the experimental magnetic field of \SI{311}{G}, and given \ch{^{15}N} P1 hyperfine couplings, we expect, based on the possible P1 Jahn-Teller axis directions and \ch{^{15}N} nuclear spin states (see Methods), transitions near \SI{935}{\mega\hertz} and \SI{954}{\mega\hertz} for the four possible $\left<111\right>$ crystallographic axes, respectively, and the nitrogen nuclear spin state $+1/2$ probed here. We observe resonances at microwave light frequencies $f_{P1}$ of \SI{934.8}{\mega\hertz} ($f_{P1,3/8}$) and \SI{953.1}{\mega\hertz} ($f_{P1,1/8}$). The subscripts indicate the fraction of the bath probed at that frequency. This confirms the presence of \ch{^{15}N} P1 centers in our sample.

We measure $T_2^*$ times for a set of eight single NV centers in the same test sample. Fig.~\ref{fig:fig4}(b) shows characteristic Ramsey interferometry data for one of these NV centers. Data is fit to an exponential decay with oscillations capturing the coupling to single nearby P1 centers, as in the Ramsey analysis in Sec.~\ref{sec:results}. While the $^{15}$NV center exhibits a $\approx$\SI{3}{\mega\hertz} splitting from its nitrogen nuclear spin, it does not contribute to NV center decoherence rate and thus is not included in CCE calculations. We are careful to drive with \SI{909}{kHz} Rabi rate pulses to avoid mixing nuclear hyperfine effects into our measurement. This NV center exhibits $T_2^*$=\SI{25\pm2}{\micro\second} (see SI Sec.~III for details of measurements). This process is repeated for eight NV centers.

Decoherence rates for the set of measured NV centers are plotted in Fig.~\ref{fig:fig4}(c) along with the calculated probability distribution function (PDF) that best fits the measured distribution as determined via MLE, discussed in the next section. The aim in the following section will be to determine which calculated distribution best fits this dataset.

\subsection{\label{subsec:mle}Maximum likelihood estimation}
Using the theoretical dependence of the coherence time distributions on thickness and P1 density, in this section we develop the maximum likelihood model (MLE) to recover the growth parameters of the given sample. Taking interpolated distributions $P(1/T_2^*)$, recovered from the numerical data, the likelihood of a given bath configuration is calculated as the joint probability of the \{$T_2^*$\} dataset for each pair of bath thickness $t$ and density $\rho$ as~\cite{Annis1953}

\begin{equation}
    \label{eq:likelihood}
    L(t,\rho)=\prod_i P(1/T_{2,i}^*|t,\rho).
\end{equation}

\begin{figure}
    \centering
    \includegraphics[width=86mm]{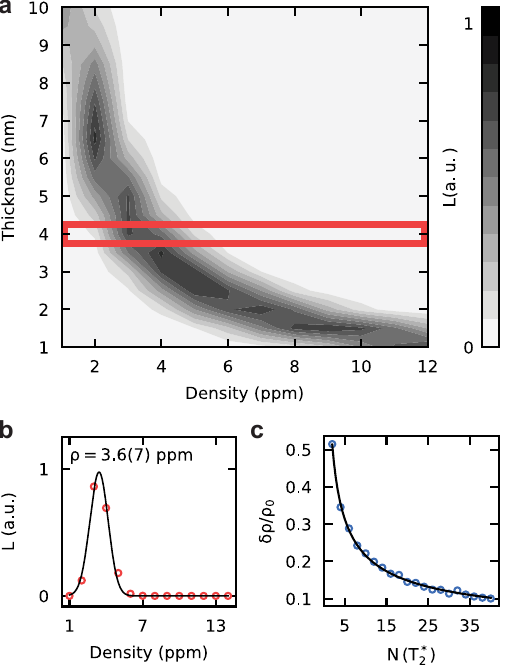}
    \caption{\bf Maximum likelihood estimation. \rm (a) Likelihood of dataset in Fig.~\ref{fig:fig4}(c) calculated for each set of bath parameters, from theoretical results. (b) Likelihood restricted to a thickness of $t_{SIMS}=\,$\SI{4}{\nano\meter} (from Fig.~\ref{fig:fig2}(e)), from which we extract a density of \SI{3.6\pm.7}{ppm}. (c)  Calculated error of density estimation across full density range with fixed thickness, calculated for 200 random datasets at each density.}
    \label{fig:fig5}
\end{figure}

The MLE procedure determines what coherence distribution best predicts the measured distribution in Fig.~\ref{fig:fig4}(c). In Fig.~\ref{fig:fig5}(a) we plot $L(t,\rho)$ over the computational phase space for the coherence times in Fig.~\ref{fig:fig4}(c). We find a band of potential bath geometries that satisfy the observed coherence time distribution, rather than uniquely predicting a single set of values. Based on the CVD growth discussed in Sec.~\ref{subsec:methods:growth}, we estimate the bath thickness at $t_{SIMS}=$\SI{4}{\nano\meter} and plot the linecut of $L$ in Fig.~\ref{fig:fig5}(b). This provides a measure of the bath density of \SI{3.6\pm.7}{ppm}, where the error is found by fitting $L(t_{SIMS},\rho)$ to a normal distribution.

We benchmark the error in the MLE procedure versus the number of coherence time samples in Fig.~\ref{fig:fig5}(c). For each number of samples, $N$, and set of bath parameters, 200 random $T_2^*$ datasets of $N$ coherence times are chosen from the numerical datasets used in Sec.~\ref{sec:results}. Then, the likelihood is calculated for a fixed thickness $t_0$, and the relative error for one dataset is calculated as $\epsilon_{\rho_0}^2=(\rho_{mle}-\rho_0)^2/\rho_0^2$, where $\rho_{mle}$ is the density such that $L(t_0,\rho_{mle})=\max\left[L(t_0,\rho)\right]$. This is averaged over a range of tested densities, plotted in Fig.~\ref{fig:fig5}(c). We calculate the error for eight samples to be \SI{25}{\percent}, corresponding to an uncertainty of \SI{0.9}{ppm} for the density estimate from Fig.~\ref{fig:fig5}(a). This is similar to the error from fitting $L$, and is stable when the thickness is varied. We fit the average error as $A\cdot N^{-p}$, shown over the calculated error in Fig.~\ref{fig:fig5}(c), finding a $N^{-1.6}$ trend.

\subsection{\label{sec:NVP1}Strong coupling yield}

\begin{figure}
    \centering
    \includegraphics[width=86mm]{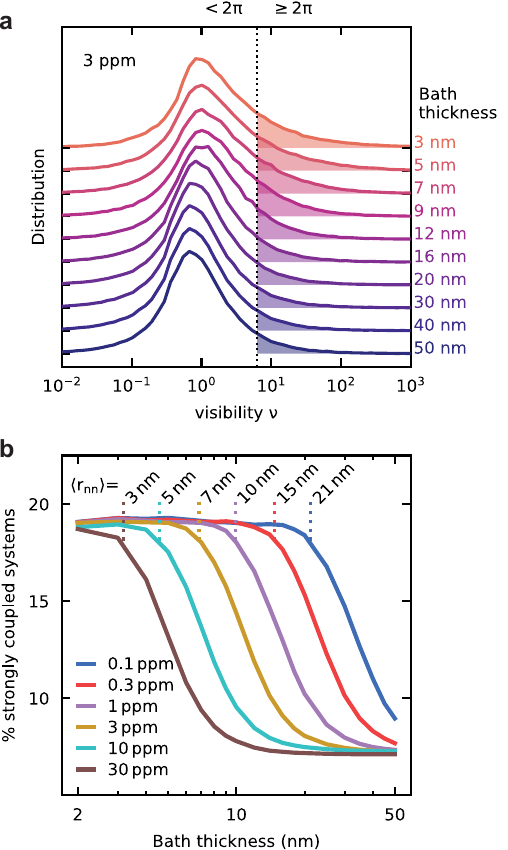}
    \caption{\bf Dimensionality dependence of strong coupling. \rm (a) Computed distribution (using PyCCE) of ratio of nearest-neighbor P1 coupling to background decoherence rate for $10^5$ \SI{3}{ppm} density P1 bath configurations with varying thickness. Curves are offset for clarity. Shaded regions right of the dashed line indicate $\nu\geq 2\pi$. (b) Percentage of NV-P1 bath systems with at least one strongly coupled bath spin for varying bath density and thickness. Average spin-spin distance is marked atop curves for each density.}
    \label{fig:fig6}
\end{figure}

Entangled qubit-based sensors promise to greatly enhance quantum sensing capabilities as compared to the current state-of-the-art~\cite{Zheng2022,Meriles2023}. The applicability of these schemes is enabled by high-yield synthesis of strongly coupled quantum systems (NV center spins and multiple single bath spins). We consider the impact of growth dimensionality on the yield of such systems analytically, quantifying our results with numerical predictions. In our calculations we consider central NV center spins and P1 center bath spins, but our approach is easily generalized to other spin systems.

Each bath spin couples to the NV center with a dipolar coupling strength $A^i_{z}$. The NV center coherence in the absence of dynamical protocols and coupled to a bath of many weakly coupled spins can be described as a product of individual coupling contributions (see Methods).

We aim to describe how likely the coupling to the nearest spin, $A_0$, is to be greater than the dephasing from the rest of the bath, $A_{\mathrm{bath}}$. The distributions of nearest neighbor distance $r_{nn}$ in two and three dimensions are
\begin{eqnarray}
    \label{eq:distributions2D}
    &&g_{2D}=\exp(-\pi r_{nn}^2\varsigma)\varsigma\,2\pi r_{nn}, \\
    \label{eq:distributions3D}
    &&g_{3D}=\exp(-4\pi r_{nn}^3\rho/3)\rho\,4\pi r_{nn}^2,
\end{eqnarray}
where $\rho$ is the 3D density and $\varsigma=\rho t$ is the 2D density for bath thickness $t$ nominally less than the average nearest neighbor distance. Notably, the distributions depend on the bath dimensionality. The bath decoherence can be estimated as follows
\begin{eqnarray}
    \Gamma_{2D}^{\mathrm{bath}}\propto \sqrt{\sum_{2D}|1/r^3|^2}&=\sqrt{\int_0^\infty dr 2\pi r\varsigma r^{-6}},\\
    \Gamma_{3D}^{\mathrm{bath}}\propto \sqrt{\sum_{3D}|1/r^3|^2}&=\sqrt{\int_0^\infty dr 4\pi r^2\rho r^{-6}}.
\end{eqnarray}
We now define the visibility $\nu$ of the nearest neighbour spin as a ratio between its coupling to the central spin $A_0$ and the decoherence rate induced by all other spins $A_\mathrm{bath}$
\begin{equation}\label{eq:visibility}
    \nu=\frac{|A_0|}{\sqrt2 A_{\mathrm{bath}}},
\end{equation}
and evaluate average $\nu$ over many bath configurations. Assuming the point dipole approximation to compute the coupling between central and bath spins, we find the average visibility at the given dimensionality as $\langle \nu_{kD}\rangle=\langle{|A_0|/\sqrt2 A_{\mathrm{bath}}}\rangle\simeq\langle{r_{nn}^{-3}/\sqrt 2\Gamma_{kD}}\rangle$, where the distributions $\Gamma_{kD}$ ($k=\{2,3\}$) are given by Eqs.~\eqref{eq:distributions2D} and \eqref{eq:distributions3D}. We note here that averaging $\Gamma_{kD}$ assumes the dephasing rate due to the rest of the spin bath follows a highly peaked distribution. We then ask if this average is larger for lower dimensional spin baths by evaluating
\begin{equation}
    \label{eq:2D3Dratio}
    \frac{\nu_{2D}}{\nu_{3D}}=\frac{\langle\sfrac{r_{nn}^{-3}}{\Gamma_{2D}}\rangle}{\langle\sfrac{r_{nn}^{-3}}{\Gamma_{3D}}\rangle}=\sqrt 2.
\end{equation}
We find that the visibility of the nearest neighbour spin is $\sqrt 2$ larger in the 2D case, pointing at the fact that yield of strongly coupled bath spins is significantly higher in the low dimensional systems. 

We confirm these analytical predictions with numerical simulations. Using the PyCCE code we generate $10^5$ \SI{50}{\nano\meter}-thick P1 electron spin baths in a (001)-oriented diamond lattice whose densities range over two orders of magnitude, and divide each bath into slices of varying thickness. For each density and thickness we compute visibility $\nu$ (Eq.~\eqref{eq:visibility}). Representative histograms for \SI{3}{ppm} spin baths are shown in Fig.~\ref{fig:fig6}(a). As the bath thickness decreases, the visibility distribution shifts to higher values, in line with the prediction from Eq.~\eqref{eq:2D3Dratio}. We follow the criterion laid out in the Methods and Eq.~\eqref{eq:strongcond} below to identify strongly coupled bath spins. We set a threshold for the visibility at $\nu\ge2\pi$. At this value coherence goes through a full oscillation period when the signal contrast reaches $1/e$.

We plot the resulting probability of obtaining strongly coupled spins in Fig.~\ref{fig:fig6}(b) for each density. At all densities, the likelihood of finding a NV-spin bath configuration with the desired coupling ratio is almost three times as high in the thin bath limit. Furthermore, there is a crossover transition for each density from three-dimensional to two-dimensional behavior, which intersects with the average nearest neighbor spacing $\langle r_{nn}\rangle=0.554\rho^{-1/3}$, obtained from Eq.~\eqref{eq:distributions3D}. Heuristically, as the thickness reduces below $\langle r_{nn}\rangle$, there are no spins proximal to the central spin in the out-of-plane direction, only in the plane of the central spin. In SI Sec.~IV we present a point of comparison between the analytical and numerical approaches, finding agreement between calculated coupling distributions and the result in Eq.~\eqref{eq:2D3Dratio}.

\section{\label{Outlook}Outlook}

Point-defects in diamond and other wide-bandgap semiconductors are promising platforms for qubit-based sensors. Deterministic synthesis of such systems will benefit from feed-forward techniques that optimize host crystal parameters for specific outcomes and applications. Additionally, such systems pave the way for entangled qubit-based sensors which hold great promise to enhance current quantum sensing capabilities. In this paper, we have demonstrated holistic quantum simulations of NV center coherence, with techniques applicable to other spin defects, as a tool for quantum system coherence characterization driven synthesis, minimizing the need for large-scale and destructive materials characterization. Practically, we showed how our approach allows for the use of rudimentary $T_2^*$ measurements to approximate \emph{in-situ} doping densities, even with little prior sample knowledge. Specifically, we have demonstrated a MLE model based on a CCE-generated distribution library as an aid to process calibration and sample characterization. This method is non-destructive and operates at the density scales relevant for quantum technologies.

Additionally, the coherence distribution results presented in this paper explore the expected sample properties in low-dimensional spin baths. By going beyond approximate analytical treatments and sampling over a wide distribution of random bath configurations with a range of central spin-bath couplings, the CCE calculations quantitatively capture the connection between bath geometries and coherence time distributions, providing an invaluable analytical tool for experimental design.

While in this work we focus on a single dominant spin bath species in low-dimensional geometries, our MLE method is not limited to this regime. CCE methods can readily be extended to additional spin bath species in diamond, as well as mixed nuclear and electronic spin baths. By calculating coherence times in these other situations, dopant densities in samples with multiple dominant noise sources can be characterized. Furthermore, the strategy presented here can be applied to other solid state hosts where qubit coherence is limited by spin bath noise.

\section{\label{methods}Methods}
Our work builds on two previously established techniques, CCE calculations~\cite{Yang2008,Yang2009} and PECVD synthesis of NV centers in diamond~\cite{Ohno2012} as described below and in Fig.~\ref{fig:fig2}. We focus on the \ch{^{15}N} isotope of nitrogen for the majority of the calculations as this allows us to experimentally distinguish intentionally doped defects from background occurring defects~\cite{Rabeau2006,Ohno2012}.

\subsection{Theory}
The dynamics of the systems are simulated using the following Hamiltonian:

\begin{equation} \label{H_e}
\begin{array}{l}
\hat H = - \gamma_e B_z \hat S_z + D\hat S_z^2 
         + \sum_i  a(m_i) \hat P_{z,i} - \gamma_e B_z \hat P_{z,i} \\
         \\
         + \sum_i \textbf{S}\textbf{A}^i \textbf{P}_i + \sum_{i \neq j} \textbf{P}_i \textbf{J}^{ij} \textbf{P}_j
         ,
\end{array}
\end{equation}
where $\gamma_e$ is the electron spin gyromagnetic ratio, $B_z$ is the magnetic field aligned with the $z$-axis, $\textbf{S}=(\hat S_x, \hat S_y, \hat S_z)$ are NV center spin operators, $D$ is the NV zero field splitting, $\textbf{P}=(\hat P_x, \hat P_y, \hat P_z)$ are spin operators of the P1 center, and $a(m_i)$ is the hyperfine coupling between the P1 \ch{^{15}N} nuclear spin and the P1 electron spin, dependent on the random orientation of the Jahn-Teller axis along one of four crystal directions and the nuclear spin state for each P1 ($m_i$), where $i$ runs over all the simulated P1 centers~\cite{Cox1994}. 
$\textbf{A}^i$ are dipolar couplings between the NV center and P1 centers, and $\textbf{J}^{ij}$ is the coupling between two P1 electron spins. The applied \SI{50}{G} is sufficiently past the high field limit and these calculations translate over to measurements at higher fields as well (see SI Sec.~IA).

 In the SI (Sec.~IB and IC) we show convergence tests for Ramsey and Hahn echo simulations versus both CCE order (1 and 4, respectively) and total number of simulated bath spins. We use CCE methods with bath state sampling~\cite{Onizhuk2021c} to achieve convergence for the electron spin bath. For each pure electron bath state the state of \ch{^{15}N} spin and the P1 crystallographic orientation is chosen at random. More details about the method are available in~\cite{Onizhuk2021b}.

The CCE approach~\cite{Yang2008,Yang2009} approximates the coherence function $\mathcal{L}(t) = \frac{\langle \sigma_-(t)\rangle}{\langle \sigma_-(0)\rangle} = \frac{\bra{0}\hat \rho (t) \ket{1}}{\bra{0}\hat \rho (0) \ket{1}}$, the normalized off-diagonal element of the density matrix $\rho_{m,n}$ of the qubit, where $m$ and $n$ are either the ground or excited spin states $\ket 0$ and $\ket 1$, respectively. $\mathcal{L}(t)$ is approximated as a product of cluster contributions:
\begin{equation}\label{gCCE_eq}
\mathcal{L}(t) = 
            \prod_{i} \tilde L^{\{i\}}
            \prod_{i, j} \tilde 
            L^{\{ij\}} ...,
\end{equation}
where $\tilde L^{\{i\}}$ is the contribution of a single bath spin, $\tilde L^{\{ij\}}$ is the contribution of spin pairs, and so on for higher order clusters (Fig.~\ref{fig:fig2}(a)). The maximum size of the cluster $n$ included in the expansion denotes the order of CCE$n$ approximation. 

The Ramsey signal is converged at the first order of CCE. As such, we can represent the high-field Ramsey coherence function in the rotating frame for a bath in a fully mixed state as~\cite{Zhao2012}: 
\begin{equation}\label{eq:cce1dephasing}
\begin{array}{l}
    \mathcal{L}(t) \approx \prod_j^N{\cos{\frac{A_{z}^jt}{2}}} \approx \exp\left[-\frac{A_{\mathrm{bath}}^2}{2} t^2\right] \prod_{i}^n {\cos{\frac{A_{z}^it}{2}}} \\
    \\
    = \exp\left[-(\frac{t}{T_2^*} )^2\right] \prod_{i}^n{\cos{\frac{A_{z}^it}{2}}} 
\end{array}
\end{equation}

where $A_{\mathrm{bath}}^2 = \frac{\sum_j (A_z^j)^2}{4}$, $T_2^*=\frac{\sqrt{2}}{A_\mathrm{bath}}$ index $i$ goes over only $n$ the strongly coupled P1 centers, and index $j$ goes over all other P1s. We define strongly coupled bath spins as those distinguishable from the background decoherence, setting threshold for its visibility (Eq.~\eqref{eq:visibility}) as:
\begin{equation}\label{eq:strongcond}
    \nu_i=\frac{\left|A_{z}^i\right|}{2}\ge 2\pi \cdot \frac{A_{\mathrm{bath}}}{\sqrt{2}},
\end{equation}
so that at least one full period of oscillation of the coherence function is visible when the signal contrast reaches $1/e$. For each random bath configuration we order the P1 spins by strength of the coupling, and one-by-one select out the strongly coupled spins until the condition (\ref{eq:strongcond}) is violated. $T_2^*$ is then recovered from the coupling to the remaining bath spins.

Ref.~\cite{Park2022} shows that CCE at second order can be used to qualitatively recover the behaviour of $T_2$ coherence times in the P1 bath. We further extend this approach, and converge CCE Hahn echo calculations at 4th order with bath-state sampling (see SI Sec.~IB and Fig.~S1(c)).

\subsection{\label{subsec:methods:growth}Materials growth}
All defects studied in this work are doped \it in-situ \rm during diamond PECVD with subsequent electron irradiation and annealing for NV center activation. This recipe constitutes our standard PECVD process for growing isotopically pure diamond with isotopically tagged NV centers, as shown in Fig.~\ref{fig:fig2}(d). High purity electronic grade (\SI{\leq10}{ppb}) diamond substrates \SI{2}{\milli\meter} by \SI{2}{\milli\meter} by \SI{0.5}{\milli\meter}, with $\langle001\rangle$ growth face and $\langle110\rangle$ sides (Element Six) were used as starting substrates. The as-received substrates were Chemical-Mechanical Polished (CMP) to a surface roughness of $R_q \SI{\leq0.4}{\pico\meter}$ by Syntek, LLC. Subsequently, these substrates were inductively coupled plasma reactive ion etched (ICP-RIE) down to remove \SI{\approx2.5}{\micro\meter} of damaged diamond surface using a composite, cycled \ch{Ar}/\ch{Cl2} and \ch{O2} plasma etching recipe. Pre-growth, the samples were annealed at \SI{1200}{\celsius} and tri-acid cleaned to mobilize/annihilate vacancy clusters and remove any amorphous/\ch{sp^2} carbon, respectively. See SI Sec.~II for a more detailed description of sample processing.

Homoepitaxial growth of diamond was performed in a custom-configured PECVD reactor~\cite{Guo2021} (Seki Diamond). The growth chamber was pumped down to \SI{8e-8}{Torr} to minimize background contamination. Thereafter, high purity \ch{H2} (\SI{99.99999}{\percent}) was introduced into the chamber, with the process microwave power of \SI{11.5}{\watt\per\milli\meter\squared} and pressure of \SI{25}{Torr} maintained throughout. The substrate temperature was maintained in the range of \SI{800\pm27}{\celsius} as tracked by an IR pyrometer. Before introduction of the diamond growth precursor, the sample was submitted to a \ch{H2} \& \ch{O2} etch (\SI{4}{\percent} of \ch{O2}) for \SI{5}{\minute} and a subsequent \SI{20}{\minute} etch using \ch{H2} only, to etch away any residual surface contaminants and defects, and expose the growth surface atomic step edges~\cite{Tallaire2008,Ohno2014b}. Thereafter, \ch{^{12}CH4} (\SI{99.9999}{\percent} chemical purity, \SI{99.99}{at.\percent} isotopic purity) is introduced as the carbon precursor. The methane-to-hydrogen ratio is maintained constant at \SI{0.1}{\percent} ($\ch{H2:^{12}CH4} = \SI{400}{sccm}:\SI{0.4}{sccm}$) as to ensure step-flow growth~\cite{Ohno2014b,Guo2021}. Growth rates for the obtained films were determined to be \SI{38\pm10}{\nano\meter\per\hour} via ex-situ secondary ion mass spectroscopy (SIMS) analysis averaged over six calibration substrates (e.g., \ch{^{12}C} overgrowth shown in Fig.~\ref{fig:fig2}(e)).

Post-growth, bulk electron irradiation with a \SI{2e14}{\centi\meter^{-2}} dose at \SI{2}{\mega eV} and a \SI{2}{\hour} anneal at \SI{850}{^{\circ}C} under an \ch{Ar} atmosphere converts a fraction of doped nitrogen into NV centers with $[\ch{^{15}NV}]\approx\SIrange{.01}{.1}{ppb}$, with the remaining nitrogen sites persisting as \ch{N_s} (P1 centers). NV activation is intentionally performed in a vacancy diffusion-limited regime~\cite{Ohno2014} in order to reliably obtain optically resolvable single NV centers. As the nitrogen doping is buried \SI{50}{\nano\meter} below the diamond surface, we do not expect band-bending effects on the defect charge states~\cite{Broadway2018,Neethirajan2023}.

\bibliography{main}


\section{Acknowledgements}
This work was primarily supported by the U.S. Department of Energy, Office of Science, Basic Energy Sciences, Materials Sciences and Engineering Division (J.C.M., F.J.H., N.D., D.D.A.) and the Design and Optimization of Synthesizable Materials with Targeted Quantum Characteristics (AFOSRFA9550-19-1-0358). We acknowledge additional support from Midwest Integrated Center for Computational Materials (MICCoM) as part of the Computational Materials Sciences Program funded by the US Department of Energy (M.O. and G. G.), the Q-NEXT Quantum Center as part of the US Department of Energy, Office of Science, National Quantum Information Science Research Centers (M.F.), and the Center for Novel Pathways to Quantum Coherence in Materials, an Energy Frontier Research Center funded by the US Department of Energy, Office of Science, Basic Energy Sciences (Y.-X. W.).  J.C.M. acknowledges prior support from the National Science Foundation Graduate Research Fellowship Program (grant no. DGE-1746045). M.O. acknowledges the support from the Google PhD Fellowship. M.W. acknowledges support from the GEM Fellowship Program.  This work made use of the Pritzker Nanofabrication Facility of the Institute for Molecular Engineering at the University of Chicago, which receives support from Soft and Hybrid Nanotechnology Experimental (SHyNE) Resource (NSF ECCS-2025633), a node of the National Science Foundation’s National Nanotechnology Coordinated Infrastructure. This work made use of the shared facilities at the University of Chicago Materials Research Science and Engineering Center, supported by the National Science Foundation under award number DMR-2011854.

\section{Author contributions}
J.C.M. and M.O. conceived the study. J.C.M. performed experimental measurements and analyzed the data. M.O. performed theoretical simulations. N.D. grew the diamond sample and analyzed SIMS results. Y.-X.W. and M.F. helped develop the MLE and strong coupling models. M. W. assisted in conducting simulations. D.D.A, G.G., F.J.H., and A.A.C. advised on all efforts. All authors contributed to writing the manuscript.

\section{Competing Interests}
The work described here is the basis of a patent application that is pending with the USPTO, filed by authors J.C.M., M.O., Y.-X.W., M.F., N.D., F.J.H., A.A.C., G.G., and D.D.A.
\end{document}